\documentclass[11pt]{article}

\newif\ifpdf
\ifx\pdfoutput\undefined
\pdffalse % we are not running PDFLaTeX
\else
\pdfoutput=1 % we are running PDFLaTeX
\pdftrue
\fi

\ifpdf
\usepackage[pdftex]{graphicx}
\else
\usepackage{graphicx}
\fi

\usepackage{cite}
\usepackage{citesupernumber}
\usepackage{nature}
\title{Links tell us about lexical and semantic Web content}

\author{Filippo Menczer\\
Department of Management Sciences\\
The University of Iowa\\
Iowa City, IA 52242\\
}

\begin{document}

\ifpdf
\DeclareGraphicsExtensions{.pdf, .png}
\else
\DeclareGraphicsExtensions{.eps, .png}
\fi

\date{}

\maketitle

\textbf{The latest generation of Web search tools is beginning to 
exploit hypertext link information to improve 
ranking\cite{Brin98,Kleinberg98} and 
crawling\cite{Menczer00,Ben-Shaul99etal,Chakrabarti99} algorithms.  The 
hidden assumption behind such approaches, a correlation between the 
graph structure of the Web and its content, has not been tested 
explicitly despite increasing research on Web 
topology\cite{Lawrence98,Albert99,Adamic99,Butler00}.  Here I 
formalize and quantitatively validate two conjectures drawing 
connections from link information to lexical and semantic Web content.  
The \emph{link-content conjecture} states that a page is similar to 
the pages that link to it, i.e., one can infer the lexical content of 
a page by looking at the pages that link to it.  I also show that 
lexical inferences based on link cues are quite heterogeneous across 
Web communities.  The \emph{link-cluster conjecture} states that pages 
about the same topic are clustered together, i.e., one can infer the 
meaning of a page by looking at its neighbours.  These results explain 
the success of the newest search technologies and open the way for 
more dynamic and scalable methods to locate information in a topic or 
user driven way.}

%:\section{Intro}

All search engines basically perform two functions: (i) crawling Web 
pages to maintain an index, and (ii) matching URLs in the index 
database against user queries.  Effective search engines achieve a 
high coverage of the Web, keep their index fresh, and rank hits in a 
way that correlates with the user's notion of relevance.  Ranking and 
crawling algorithms use cues from words and hyperlinks, associated 
respectively with \emph{lexical} and \emph{link topology}.  In the 
former, two pages are close to each other if they have similar textual 
content; in the latter, if there is a short path between them.  Lexical 
metrics are traditionally used by search engines to rank hits 
according to their similarity to the query, thus attempting to infer 
the semantics of pages from their lexical representation.  Similarity 
metrics are derived from the vector space model\cite{Salton83}, that 
represents each document or query by a vector with one dimension for 
each term and a weight along that dimension that estimates the term's 
contribution to the meaning of the document.  The \emph{cluster 
hypothesis} behind this model is that a document lexically close to a 
relevant document is also relevant with high probability\cite{vanR79cluster}.  
Links have traditionally been used by search engine crawlers only in 
exhaustive, centralized algorithms.  However the latest generation of 
Web search tools is beginning to integrate lexical and link metrics to 
improve ranking and crawling performance through better models of 
relevance.  The best known example is the \emph{PageRank} metric used 
by Google: pages containing the query's lexical features are ranked 
using query-independent link analysis\cite{Brin98}.  Links are also 
used in conjunction with text to identify hub and authority pages for 
a certain subject\cite{Kleinberg98}, determine the reputation of a 
given site\cite{Mendelzon00}, and guide search agents crawling on 
behalf of users or topical search 
engines\cite{Menczer00,Ben-Shaul99etal,Chakrabarti99}.  

%:\section{The link-content conjecture}

To study the connection between link and lexical topologies, I 
conjecture a positive correlation between distance measures defined in 
the two spaces.  Given any pair of Web 
pages $(p_{1},p_{2})$ we have well-defined distance functions 
$\delta_{l}$ and $\delta_{t}$ in link and lexical space, 
respectively.  To compute $\delta_{l}(p_{1},p_{2})$ we use the Web 
hypertext structure to find the length, in links, of the shortest path 
from $p_{1}$ to $p_{2}$. (This is not a metric distance 
because it is not symmetric in a directed graph, but 
for convenience I refer to $\delta_{l}$ as ``distance''.) 
To compute $\delta_{t}(p_{1},p_{2})$ we 
can use the vector representations of the two pages, where the vector 
components (weights) of page $p$, $w_{p}^{k}$, are computed for terms 
$k$ in the textual content of $p$ given some weighting scheme.  One 
possibility would be to use Euclidean distance in this word vector 
space, or any other $L_{z}$ norm.
However, $L_{z}$ metrics have a dependency on the 
dimensionality of the pages, i.e., larger documents tend to appear 
more distant from each other than shorter ones, irrespective of 
content.  To circumvent this problem, one can instead define a metric 
based on the \emph{similarity} between pages.
Let us use the \emph{cosine similarity} function, a standard measure 
in information retrieval:
\begin{equation}
	\sigma(p_{1},p_{2}) = \frac{\sum_{k \in p_{1} \cap p_{2}} 
	w_{p_{1}}^{k} w_{p_{2}}^{k}}
	{\sqrt{\sum_{k \in p_{1}} (w_{p_{1}}^{k})^{2} 
	\sum_{k \in p_{2}} (w_{p_{2}}^{k})^{2}}}.
	\label{eq:sim}
\end{equation}

\begin{figure}[hp]
	\centering
	\includegraphics{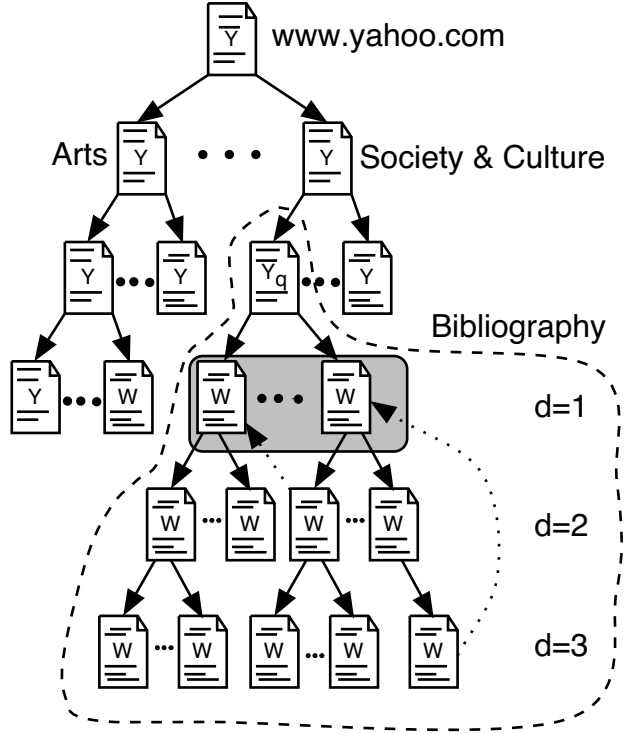} 
	\caption{Representation of the data collection.
	100 topic pages were chosen in the Yahoo directory owing to 
	this portal's wide popularity. Yahoo category
	pages are marked ``Y'', external pages are marked ``W''.
	The topic pages were chosen among ``leaf'' categories, i.e.  
	without sub-categories.  This way 
	the external pages linked by a topic page (``Yq'') represent the
	relevant set compiled for that topic by the Yahoo editors (shaded).
	Topics were selected in breadth-first order and therefore 
	covered the full spectrum of Yahoo top-level categories.  
	In this example the topic is \texttt{SOCIETY CULTURE BIBLIOGRAPHY}.
	Arrows represent hyperlinks and dotted arrows are examples of links
	pointing back to the relevant set.
	For each topic, we performed 
	a breadth-first crawl up to a depth of 3 links.  The crawl set is 
	represented inside the dashed line.
	To obtain meaningful and comparable 
	statistics at $\delta_{l}=1$, only topic pages with at least 
	5 external links were used, and only the first 10 links 
	for topic pages with over 10 links.  Each crawl 
	was stopped if 10,000 pages had been downloaded at depth $\delta_{l}=3$
	from the start page.  A timeout of 60 seconds was applied for each 
	page.  The resulting collection comprised 376,483 pages.  The text of 
	each fetched page was parsed to extract links and terms. Terms were 
	conflated using a standard stemming algorithm\cite{Porter80}.
	A common TFIDF weighting scheme\cite{Jones72} was employed to 
	represent each page in word vector space.  This model assumes a 
	global measure of term frequency across pages (inverse document 
	frequency).  To make the measures scalable with the maximum crawl 
	depth (a parameter), inverse document frequency was computed as a 
	function of distance from the start page, among the set of 
	documents within that distance from the source.  Formally, for 
	each topic $q$, page $p$, term $k$ and depth $d$:
	$w_{p,d,q}^{k} = tf(k,p) \cdot idf(k,d,q)$
	where $tf(k,p)$ is the number of occurrences 
	of term $k$ in page $p$ and
	$idf(k,d,q) = 1 + \ln\left(\frac{N_{d}^{q}}{N_{d}^{q}(k)}\right)$.
	Here $N_{d}^{q}$ is the size of the cumulative page set 
	$P_{d}^{q} = \{ p : \delta_{l}(q,p) \leq d \}$, and
	$N_{d}^{q}(k)$ is the size of the subset of $P_{d}^{q}$ of pages
	containing term $k$.}
	\label{yahoo}
\end{figure}

According to the link-content conjecture, $\sigma$ is anticorrelated 
with $\delta_{l}$.  The idea is to measure the correlation between the 
two distance measures across pairs of pages.  Figure~\ref{yahoo}
illustrates how a collection of Web pages was crawled and processed
for this purpose.

%:\subsection{Correlation of lexical and link distance}

The link distances $\delta_{l}(q,p)$ and similarities $\sigma(q,p)$ 
were averaged for each topic $q$ over all pages $p$ in the crawl set 
$P_{d}^{q}$ for each depth $d$:
\begin{eqnarray}
	\delta(q,d) &\equiv& \langle \delta_{l}(q,p) \rangle_{P_{d}^{q}} =
		\frac{1}{N_{d}^{q}} \sum_{i=1}^{d} i \cdot 
		(N_{i}^{q} - N_{i-1}^{q}) \label{eq:Laver} \\
	\sigma(q,d) &\equiv& \langle \sigma(q,p) \rangle_{P_{d}^{q}} = 
		\frac{1}{N_{d}^{q}} \sum_{p \in P_{d}^{q}} 
		\sigma(q,p). \label{eq:Saver}
\end{eqnarray}

\begin{figure}[hp]
	\centering
	\includegraphics{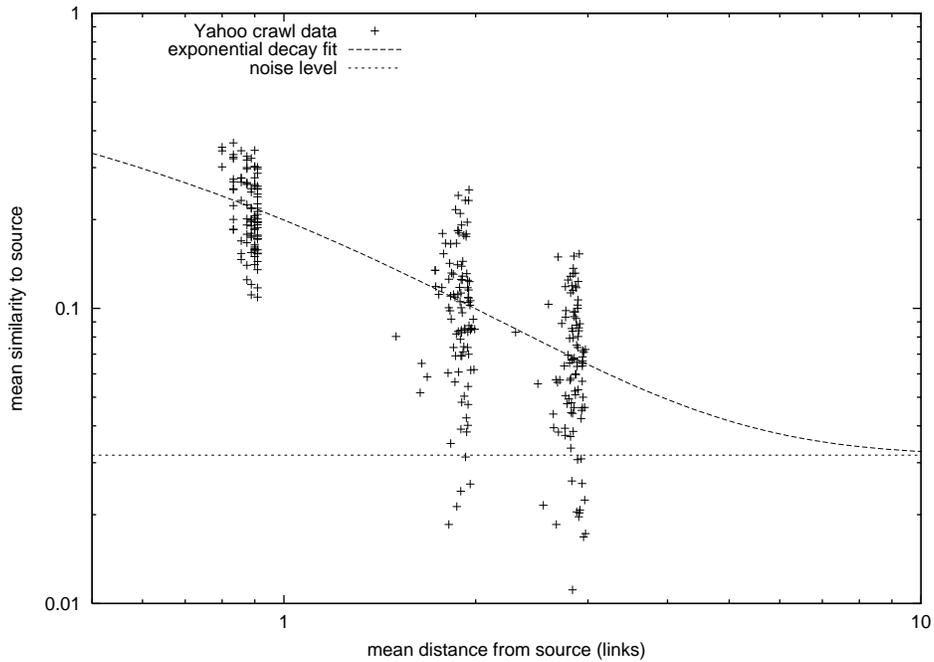} 
	\caption{Scatter plot of $\sigma(q,d)$ versus 
	$\delta(q,d)$ for topics $q=0,\ldots,99$ and
	depths $d=1,2,3$. Pearson's correlation coefficient $\rho = -0.76,
	p<0.0001$. The similarity 
	noise level $\sigma_{\infty}$ and an exponential decay fit 
	of the data and are also shown. 
	$\sigma_{\infty}$ was computed by comparing each topic 
	page to external pages linked from different Yahoo categories:
	$\sigma_{\infty} \equiv \left\langle 
	\frac{1}{N_{1}^{q'}} \sum_{p \in P_{1}^{q'}} \sigma(q,p) 
	\right\rangle_{\{q,q': q \neq q'\}} 
	\approx 0.0318 \pm 0.0006$.
	The regression yielded 
	parametric estimates $\alpha_{1} \approx 1.8$ and $\alpha_{2} \approx 
	0.6$.}
	\label{scatter}
\end{figure}

The 300 measures of $\delta(q,d)$ and $\sigma(q,d)$ 
from Equations~\ref{eq:Laver} and \ref{eq:Saver} are shown in  
Figure~\ref{scatter}.  The two metrics are indeed well 
anticorrelated and predictive of each other with high 
statistical significance.  This quantitatively 
confirms the link-content conjecture.

%:\subsection{Range of link-based lexical predictions}

To analyze the decrease in the reliability of lexical content 
inferences with distance from the topic page in link space one can 
perform a nonlinear least-squares fit of these data to a family of 
exponential decay models:
\begin{equation}
	\sigma(\delta) \sim \sigma_{\infty} + 
		(1 - \sigma_{\infty}) e^{-\alpha_{1} \delta^{\alpha_{2}}}
	\label{sim-decay}
\end{equation}
using the 300 points as independent samples. Here $\sigma_{\infty}$ 
is the noise level in similarity.
Note that while starting from Yahoo pages may bias $\sigma(\delta<1)$ 
upward, the decay fit is most affected by the constraint 
$\sigma(\delta=0) = 1$ (by definition of similarity) and by the 
longer-range measures $\sigma(\delta>1)$.  
The similarity decay fit curve is also shown in Figure~\ref{scatter}. It
provides us with a rough estimate of how far in link space one 
can make inferences about lexical content.

%:\subsection{Heterogeneity of link-based lexical cues}

\begin{figure}[hp]
\centering
\begin{tabular}{rcrc}
	\textbf{a}
	&
	\multicolumn{3}{c}{\raisebox{-1.75in}{\includegraphics{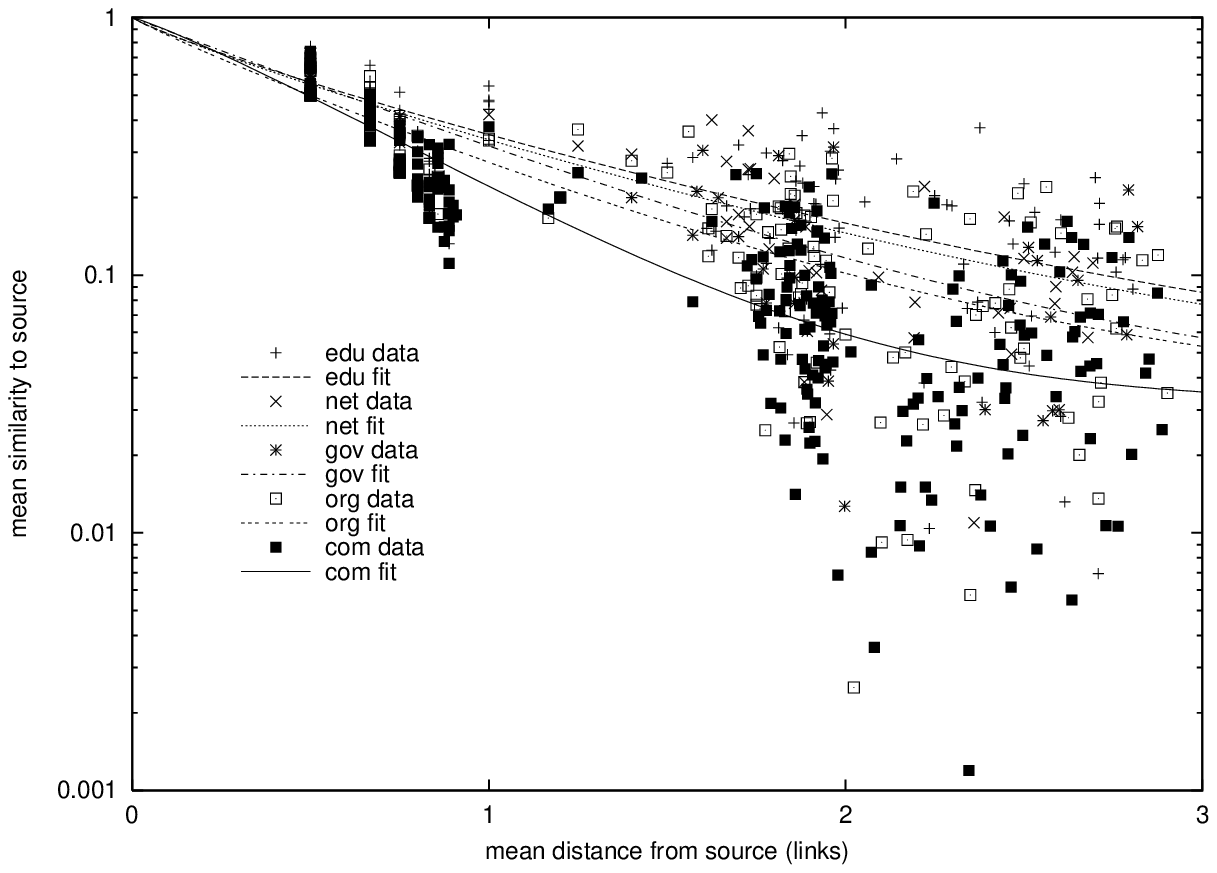}}} \\
	\textbf{b}
	&
	\begin{tabular}{ccc}
		Domain & $\alpha_{1}$ & $\alpha_{2}$ \\
		\hline
		\texttt{edu} & $1.11 \pm 0.03$ & $0.87 \pm 0.05$ \\
		\texttt{net} & $1.16 \pm 0.04$ & $0.88 \pm 0.05$ \\
		\texttt{gov} & $1.22 \pm 0.07$ & $1.00 \pm 0.09$ \\
		\texttt{org} & $1.38 \pm 0.03$ & $0.93 \pm 0.05$ \\
		\texttt{com} & $1.63 \pm 0.04$ & $1.13 \pm 0.05$ \\
		\hline
	\end{tabular}
	& 
	\textbf{c}
	&
	\raisebox{-0.8in}{\includegraphics[width=2.5in]{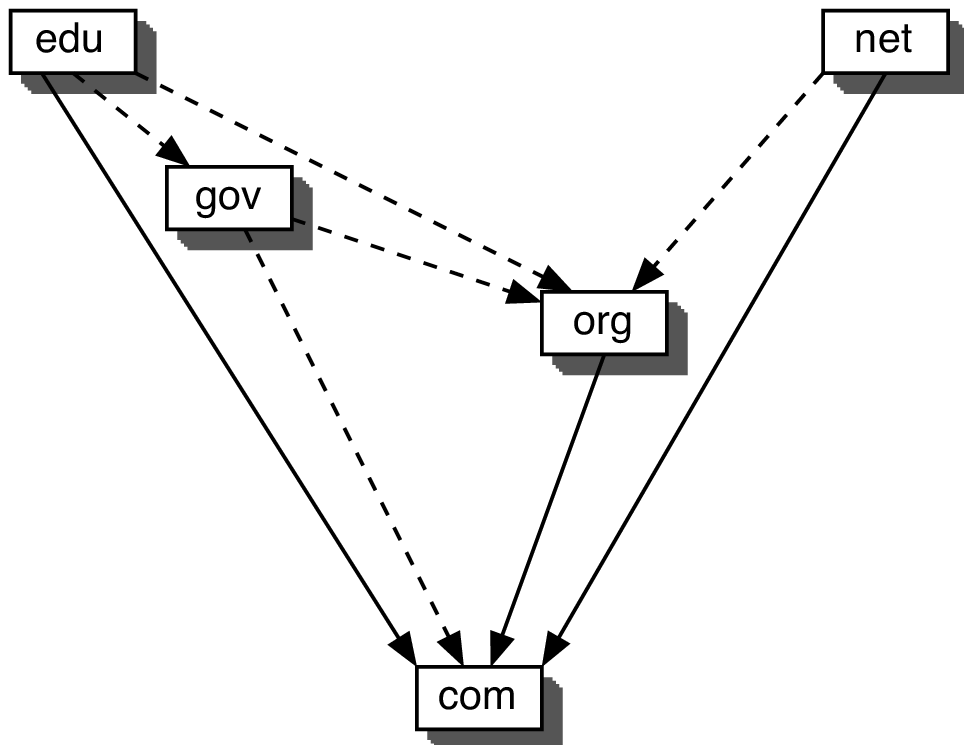}}
\end{tabular}
\caption{\textbf{a.} Scatter plot of $\sigma(q,d)$ versus 
$\delta(q,d)$ for topics $q=0,\ldots,99$ and
depths $d=1,2,3$, for each of the major US top-level domains.
The domain sets were obtained by simulating crawlers that only 
follow links to servers within each domain.  
An exponential decay fit is also shown for each domain.
\textbf{b.} Exponential decay model parameters obtained by 
nonlinear least-squares fit of each domain data. 
\textbf{c.} Summary of statistically 
significant differences (at the 68.3\% confidence level) between the 
parametric estimates; dashed arrows represent significant differences 
in $\alpha_{1}$ only, and solid arrows significant differences in 
both $\alpha_{1}$ and $\alpha_{2}$.}
\label{domains}
\end{figure}

How heterogeneous is the reliability of lexical inferences based on 
link neighbourhood across communities of Web content providers?  To 
answer this question the crawled pages were divided up into connected 
sets within top-level Internet domains.  The scatter plot of the 
$\delta(q,d)$ and $\sigma(q,d)$ measures for these domain-based crawls 
is shown in Figure~\ref{domains}a.  The plot illustrates the 
heterogeneity in the reliability of lexical inferences based on link 
cues across domains.  The parameters obtained from fitting each domain 
data to the exponential decay model of Equation~\ref{sim-decay} 
(Figure~\ref{domains}b) estimate how reliably links point to lexically 
related pages in each domain.  A summary of the statistically 
significant differences among the parametric estimates is shown in 
Figure~\ref{domains}c.  It is evident that, for example, 
academic Web pages are better connected to each other than commercial 
pages in that they do a better job at pointing to other similar pages.  
In other words it is easier to find related pages browsing through 
academic pages than through commercial pages.  This is not surprising 
considering the different goals of the two communities.

%:\section{The link-cluster conjecture}

The link-cluster conjecture is a link-based analog of the cluster 
hypothesis, stating that pages within a few links from a relevant 
source are also relevant with high probability.
Here I experimentally assess the extent to which relevance is 
preserved within link space neighbourhoods, and the decay in expected 
relevance as one browses away from a relevant page.  

The link-cluster conjecture has been implied or stated in various
forms\cite{Kleinberg98,Gibson98,Brin98,Chakrabarti98etal,Dean99,Davison00}. 
One can most simply and generally state it in terms of the conditional 
probability that a page $p$ is relevant with respect to some query $q$, 
given that page $r$ is relevant and that $p$ is within $d$ links from $r$:
\begin{equation}
	R_q(d) \equiv 
		\Pr[rel_q(p) \: | \: rel_q(r) \wedge \delta_{l}(r,p) \leq d]
\end{equation}
where $rel_q()$ is a binary relevance assessment with respect to $q$. 
In other words a page has a higher than random probability of being 
about a certain topic if it is in the neighbourhood of other pages 
about that topic. $R_q(d)$ is the posterior relevance probability 
given the evidence of a relevant page nearby. The simplest form of 
the link-cluster conjecture is stated by comparing $R_q(1)$ to the 
prior relevance probability $G_q$:
\begin{equation}
	G_q \equiv \Pr[rel_q(p)]
\end{equation}
also known as the \emph{generality} of the query. If link 
neighbourhoods allow for semantic inferences, then the following 
condition must hold:
\begin{equation}
	\lambda(q,d=1) \equiv \frac{R_q(1)}{G_q} > 1.
	\label{def_L}
\end{equation}
To illustrate the meaning of the link-cluster conjecture, consider 
a random crawler (or user) searching for pages about a topic $q$. 
Call $\eta_q(t)$ the probability that the crawler hits a relevant 
page at time $t$. Solving the recursion
\begin{equation}
	\eta_q(t+1) = \eta_q(t) \cdot R_q(1) + (1 - \eta_q(t)) \cdot G_q
\end{equation}
for $\eta_q(t+1) = \eta_q(t)$ yields the stationary hit rate 
\begin{equation}
	\eta_q^* = \frac{G_q}{1 + G_q - R_q(1)}.
\end{equation}
The link-cluster conjecture is a necessary and sufficient condition 
for such a crawler to have a better than chance hit rate, thus 
justifying the crawling (and browsing!) activity:
\begin{equation}
	\eta_q^* > G_q \Longleftrightarrow \lambda(q,1) > 1.
\end{equation}

Definition~\ref{def_L} can be generalized to likelihood factors over 
larger neighbourhoods:
\begin{equation}	
	\lambda(q,d) \equiv \frac{R_q(d)}{G_q} \stackrel{d \rightarrow 		
\infty}{\longrightarrow} 1
	\label{L-def}
\end{equation}
and a stronger version of the conjecture can be formulated as follows:
\begin{equation}
	\lambda(q,d) \gg 1 \; \mbox{for} \; \delta(q,d) < \delta^*
\end{equation}
where $\delta^*$ is a critical link distance beyond which semantic 
inferences are unreliable.

%:\subsection{Semantic clusters in link neighbourhoods}

I first attempted to measure the likelihood factor $\lambda(q,1)$ for 
a few queries and found that 
\linebreak $\langle \lambda(q,1) \rangle_q \gg 1$,
but those estimates were based on very noisy relevance 
assessments\cite{Menczer97b}. To obtain a reliable quantitative 
validation of the stronger link-cluster conjecture, I repeated such 
measurements on the data set described in Figure~\ref{yahoo}. 

\begin{figure}[hp]
	\centering
	\includegraphics{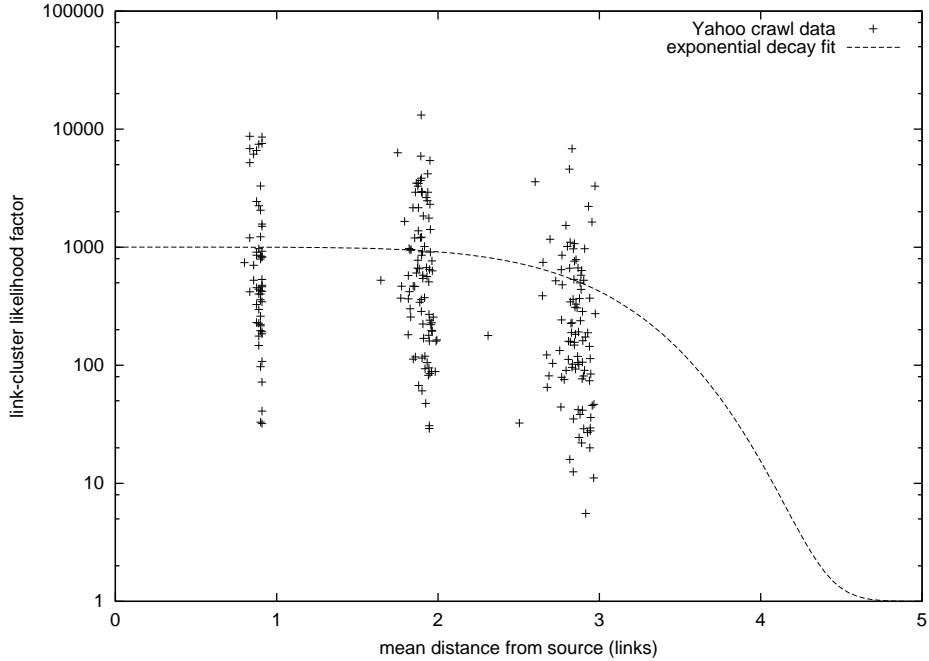}
	\caption{Scatter plot of $\lambda(q,d)$ versus 
	$\delta(q,d)$ for topics $q=0,\ldots,99$ and depths $d=1,2,3$.
	Pearson's $\rho = -0.1, p=0.09$. In computing $\lambda(q,d)$ 
	from Definition~\ref{L-def}, the relevant set $Q_q$ compiled by the 
	Yahoo editors for each topic $q$ was used to estimate
	$R_q(d) \simeq \frac{|P_{d}^{q} \cap Q_q|}{N_{d}^{q}}$
	(cf. dotted links in Figure~\ref{yahoo}). 
	Generality was approximated by
	$G_q \simeq \frac{|Q'_q|}{|\bigcup_{q' \in Y} Q'_{q'}|}$ where
	all of the relevant links for each topic $q$ are included in $Q'_q$, 
	even for topics where only the first 10 links were used in
	the crawl ($Q'_q \supseteq Q_q$), and 
	the set $Y$ in the denominator includes all Yahoo leaf categories. 
	An exponential decay fit of the data is
	also shown. The regression yielded parametric estimates 
	$\alpha_{3} \approx 1000$, $\alpha_{4} \approx 0.002$ and $\alpha_{5} 
	\approx 5.5$.}
	\label{likelihood}
\end{figure}

The 300 measures of $\lambda(q,d)$ thus obtained are plotted versus 
$\delta(q,d)$ from  Equation~\ref{eq:Laver} in  
Figure~\ref{likelihood}. Closeness to a relevant page in link space 
is highly predictive of relevance, increasing the relevance 
probability by a likelihood factor $\lambda(q,d) \gg 1$ over the 
range of observed distances and queries.

%:\subsection{Expected relevance decay in link space}

We also performed a nonlinear least-squares fit of these data to a 
family of exponential decay functions using the 300 points as 
independent samples: 
\begin{equation}
	\lambda(\delta) \sim 1 + \alpha_3 e^{-\alpha_4 \delta^{\alpha_5}}.
\end{equation}
Note that this three-parameter model is more complex than the one in 
Equation~\ref{sim-decay} because $\lambda(\delta=0)$ must also be 
estimated from the data ($\lambda(q,0) = 1/G_q$). The 
relationship between link distance and the semantic likelihood factor 
is less regular than between link distance and lexical similarity. 
The resulting fit (also shown in 
Figure~\ref{likelihood}) provides us with a rough estimate of how 
far in link space we can make inferences about the semantics 
(relevance) of pages, i.e., up to a critical distance $\delta^*$ 
between 4 and 5 links.

%:\section{Discussion}

It is surprising that the link-content and link-cluster conjectures 
have not been formalized and addressed explicitly before, especially 
when one looks at the considerable attention recently received by the 
Web's graph topology\cite{Lawrence98,Butler00}.  The correlation 
between Web links and content takes on additional significance in 
light of link analysis studies that tell us the Web is a ``small 
world'' network, i.e., a graph with an inverse power law distribution 
of in-links and out-links\cite{Albert99,Adamic99}.  Small world 
networks have a mixture of non-random local structure and non-local 
random links. Such a topology creates short paths between pages, whose length scales logarithmically with the number of Web pages.  The present 
results indicate that the Web's local structure is created by the 
semantic clusters resulting from authors linking their pages to 
related resources.

The link-cluster and link-content conjectures have important normative 
implications for future Web search technology.  For example the 
measurements in this paper suggest that topic driven crawlers should 
keep track of their position with a bias to remain within a few links 
from some relevant source.  In such a range hyperlinks create 
detectable signals about lexical and semantic content, despite the 
Web's chaotic lack of structure.  Absent such signals, the short paths 
predicted by the small world model might be very hard to locate for 
localized algorithms \cite{Kleinberg00}.  In general the present 
findings should foster the design of better search tools by 
integrating traditional search engines with topic- and query-driven 
crawlers\cite{Menczer01} guided by \emph{local} link and lexical 
clues.  Smart crawlers of this kind are already emerging (see for 
example \texttt{http://myspiders.biz.uiowa.edu}).  
Due to the size and dynamic nature of the Web, the 
efficiency-motivated search engine practice of keeping query 
processing separate from crawling leads to poor trade-offs between 
coverage and recency\cite{Lawrence99}.  Closing the loop from user 
queries to smart crawlers will lead to dynamic indices 
with more scalable and user-driven update algorithms than 
the centralized ones used today.

\subsection*{Acknowledgements}

{\small The author is grateful to D. Eichmann, P. Srinivasan, W.N. 
Street, A.M. Segre, R.K. Belew and A. Monge for helpful comments 
and discussions, and to M. Lee and M. Porter for contributions 
to the crawling and parsing code.}

\textbf{Correspondence and requests for materials should be sent to 
the author (email: \linebreak \texttt{filippo-menczer@uiowa.edu}).}

\newpage

\newpage

\newpage

\newpage

\end{document}
\end